\documentstyle[draft]{mn}
\begin{document}
\input{epsf.sty}
\title[CO emission from discs around isolated HAeBe and Vega-excess stars]
{CO emission from discs around isolated HAeBe and Vega-excess stars}
\author[W. R. F. Dent, J. Greaves, I. M. Coulson]
       {W. R. F. Dent$^1$, J. S. Greaves$^1$, I. M. Coulson$^2$\\
 $^1$UK Astronomy Technology Centre, Royal Observatory, 
Blackford Hill, Edinburgh EH9 3HJ, Scotland\\
 $^2$Joint Astronomy Centre, 660 N. A'ohoku Place, Hilo, Hawaii 96720,
USA\\
}


\pagerange{000--000}
%
%
%
\def\cm{\,{\rm cm}}
\def\cc{\,{\rm cm^{-3}}}
\def\asec{\,{\rm ''}}
\def\amin{\,{\rm '}}
\def\kps{\,{\rm {km~s^{-1}}}}
\def\Msun{\,{\rm M_{\odot}}}
\def\Lsun{\,{\rm L_{\odot}}}
\newcommand{\degree}{\mbox{\,$^\circ$}}        
\newcommand{\micron}{\mbox{\,${\mu}$m}}        
\newcommand{\elec}{\mbox{\,e$^{-}$}}           
\def\elec{e$^{-}$}
%
\def\aa{{A\&A}}
\def\aar{{A\&ARev}}
\def\aas{{A\&ASup}}
\def\aj{{AJ}}
\def\apj{{ApJ}}
\def\apjl{{ApJLet}}
\def\apjs{{ApJSup}}
\def\araa{{ARAA}}
\def\ass{{A\&Sp.Sci.}}
\def\mnras{{MNRAS}}
\def\pasp{{PASP}}
\setlength{\topmargin}{-10mm}
\maketitle
\begin{abstract}

We describe results from a survey for J=3-2 $^{12}$CO emission from
visible stars classified as having an infrared excess.
The line is clearly detected in 21 objects, and significant molecular
gas ($\geq 10^{-3}$ Jupiter masses) is found to be
common in targets with infrared excesses
$\geq$0.01 ($\geq$56 per cent of objects), but rare for those with smaller
excesses ($\sim$10 per cent of objects).

A simple geometrical argument based on the infrared
excess implies that disc opening angles are typically
$\geq 12\degree$ for objects with detected
CO; within this angle, the disc is
optically thick to stellar radiation and shields the
CO from photodissociation. Two or three CO discs have an unusually
low infrared excess ($\leq$0.01), implying the shielding disc is physically
very thin ($\leq 1\degree$).

Around 50 per cent of the detected line profiles are double-peaked,
while many of the rest have significantly broadened lines,
attributed to discs in Keplerian rotation.
Simple model fits to the line
profiles indicate outer radii in the range 30-300 au, larger
than found through fitting continuum SEDs, but
similar to the sizes of debris discs around main sequence stars.
As many as 5 have outer radii smaller than the Solar System (50 au),
with a further 4 showing evidence of gas in the disc at radii smaller
than 20~au.
The outer disc radius is independent of the stellar spectral type
(from K through to B9), but there is evidence of
a correlation between radius and total dust mass. Also the mean disc size
appears to decrease with time: discs around stars of
age 3-7 Myr have a mean radius 
$\sim$210 au, whereas discs of age 7-20 Myr are a factor of 3 smaller.
This shows that a significant mass of gas (at least
$2 \mathrm{M_{\oplus}}$)
exists beyond the region of planet formation for up to $\sim$7 Myr, and
may remain for a further $\sim$10 Myr within this region.

The only bona fide debris disc with detected CO is HD9672; this shows
a double peaked CO profile and is the most compact gas disc observed,
with a modelled outer radius of 17 au.
In the case of HD141569, detailed modelling of the line profile
indicates gas may lie in two rings, with radii of 90 and 250 au,
similar to the dust structure seen in 
scattered light and the mid-infrared. In both AB Aur and HD163296 we
also find the sizes of the molecular disc and dust scattering disc
are similar; this suggests that the molecular gas and small dust
grains are closely co-located.

\end{abstract}

\vskip 5mm

\vskip 10mm

\section{Introduction}

Vega-excess stars were initially identified as being apparently
normal main-sequence stars with an excess
of emission above that of the photosphere at $\lambda \geq 12 \micron$
(Aumann et al., 1984).
Several lists of these objects have been compiled, such as that
of Mannings \& Barlow (1998), who cross-compared the
IRAS Faint Source and the Michigan
Spectral Catalogues. They found a total of $\sim$ 110 stars with
detectable excesses, of which around 60 per cent are of spectral type B or A.
Their Spectral Energy Distributions (SEDs)
are very similar to many objects at the later stages of pre-main-sequence
evolution, such as Herbig AeBe stars. HAeBe stars have spectral
types later than B8 with clear optical emission lines and nearby nebulosity,
both indicative of youth (Herbig, 1960). They are assumed to be the
high-mass equivalents of T Tauri stars.
Th\'e et al. (1994) identified $\sim$115 bona fide HAeBe
candidates, and extended the spectral range to Fe stars.
Many of the classical HAeBe stars are associated with nearby star formation,
such as ambient molecular gas, optical nebulosity, outflows or
infalling envelopes; typically this is found at radii of
$\sim 10^3 - 10^6$au, and so can confuse studies of the stars themselves.
However, Grinen et al. (1991), Meeus et al. (2001) and others identified
a sub-class of these stars which are relatively isolated, having
no nearby cloud and a low line-of-sight extinction.
The absence of nearby active star formation
suggests that they are older than their classical brethren. Comparison with
stellar evolutionary tracks
give ages of a few Myr (e.g. van den Ancker et al., 1998), as
compared with $10^4 - 10^5$yr for classical HAeBe stars (eg
Fuente et al., 2002; Greaves et al., in prep.).
Observations of these isolated stars are easier to interpret as their
circumstellar disc emission is less confused by the
ambient gas envelope (e.g. Dominik et al., 2003).

Because of the similarity of the SEDs, there is some confusion in
the classification of Vega-excess and HAeBe stars. Dunkin et
al. (1997) and Coulson et al. (1998) suggested some of the ``dusty''
Vega-excess objects are actually at an intermediate stage
between HAeBe and main-sequence Vega-type stars.
This class of isolated HAeBe star or dusty Vega-excess object may be
therefore lie at an important evolutionary stage
at the end of the pre-main-sequence phase, and
are therefore sometimes known as ``transition'' objects
(e.g. Malfait et al., 1998).

The material around HAeBe and Vega-excess stars is most commonly
studied in dust continuum emission. However, molecular gas - traced
by the low-J lines of CO - has also been detected in a small
number of objects (e.g. Zuckerman et al., 1995; Coulson et al., 1998;
Greaves et al., 2000; Thi et al., 2001).
In many cases the signal:noise has been too low to obtain details
of the line profiles, although some show a double-peaked line.
In five of the brightest objects,
the line emission has been resolved using mm-wave interferometry, revealing
that the gas is located in a rotating disc (Manning \& Sargent, 1997, 2000;
Pi\'etu et al., 2003). A disc would also
explain the simultaneous high fractional excess emission,
and low visual extinction (e.g. van den Ancker et al., 1998).
Attempts to detect mm-wave CO emission
from the archetypal Vega-excess stars such as Vega and
$\beta$ Pictoris have proved unsuccessful, with measured
CO/dust depletions of more than 10$^3$ (Dent et al., 1995; Liseau, 1999).
This is likely to be caused by photodissociation from stellar UV;
however if there remain some regions of the disc where
the optical extinction to the star and interstellar radiation field
is $\geq 3$ mag, significant CO could still be present
(e.g. Greaves et al., 2000; Kamp et al., 2003).

The following observations comprise a more general survey
of infrared excess stars for molecular gas, as traced
by the J=3-2 transition of $^{12}$CO. Targets are identified as
either Vega-excess or isolated HAeBe stars.
We have then compared the line profiles with a simple
disc model to determine some of the basic system parameters.

\section{Observations}

The data were taken using the facility heterodyne receiver RxB3
on the JCMT. Observations were made at the
$^{12}$CO J=3-2 line rest frequency of 345.796 GHz,
using the secondary chopper to obtain a sky reference
150 arcsec distant in azimuth. Spectra from both orthogonally-polarised
mixers were averaged together. Total integration
times were 30 or 60 minutes, although a few selected
stars were targeted for longer integrations (up to 8 hours)
to better delineate the line shapes. Most data were obtained as part
of a poor-weather backup program carried out between 2000 September
and 2001 January, and 2003 January to April. A few longer integration
observations of selected targets were carried out in 2004 June.
Additional data were also
obtained from the JCMT archive, a subset of which has previously
been published (Coulson et al., 1998; Greaves et al., 2000; Thi et al.,
2001). In these cases, we have coadded the existing JCMT data with
the new results in order to increase the effective integration time.

The target list was taken from the following surveys:

\begin{itemize}

\item The list of Vega-like systems with bright IRAS excesses from
Sylvester et al. (1996) (17 out of their 24 stars).

\item Young (1-10 Myr) stars with associated dust from Zuckerman et al.
(1995) (13 out of 16 stars).

\item Vega-excess stars with the largest far-infrared flux from
Mannings \& Barlow (1998).

\item Isolated HAeBe stars from Malfait et al. (1998), van den Ancker
et al. (1998) and Grady et al. (1996).

\end{itemize}

As mentioned above, in many cases the simple identification of a star as
luminosity Class V with an IRAS excess means that the target list
includes both isolated Herbig HAeBe or even T Tauri stars, as well
as Vega-excess stars. We have avoided the classical embedded HAeBe stars.
In addition, a small number of stars have subsequently been identified 
being luminosity Class III or IV, or have revised distance estimates; these
have been identified in the analysis.

The basic target information and results are
given in Table~1; distances in most cases
are from the Hipparcos catalogue and when not available we have used
literature values or those derived from the visual magnitude.
Also listed are the fractional excess luminosities, $f$, 
ie the luminosity radiated from the system in excess of the photosphere,
as a fraction of the total stellar photospheric emission.  
In most cases $f$ is
obtained from the literature, or by fitting modified black-bodies
to the available optical through to sub-mm fluxes. This fitting
assumes in most cases a single temperature for the dust,
with an opacity index, $\beta = 1.0$ (e.g. Dent et al., 2000).
The typical uncertainty in the fractional excess is estimated to be
a factor of two, mostly due to the incomplete wavelength coverage.

The CO integrated emission or 1-$\sigma$ upper limits are given in the Table.
In the case of a line detection, the centroid CO velocity and
the stellar velocity from the literature are also listed.
In most cases these are in reasonable agreement, giving
confidence that the observed
gas is associated with the star and not with background Galactic emission.
Despite the targetting of isolated objects, 3 or 4 objects showed bright,
narrow positive or negative lines, indicating an extended ambient cloud
either associated with the star or along the line of sight.

\begin{table*}
\begin{center}
\caption{\sf Target stars and observational results.}
\begin{tabular}[t]{llllllllll} \hline\
HD & Other & Sp. & d$^{(2)}$ & $f^{(3)}$ & CO intensity$^{(4)}$ & v$_{CO}$ & v$_*$ & Notes$^{(7)}$ \& references\\
   & name  & type$^{(1)}$ & (pc)      &     & (K km s$^{-1}$) & (km s$^{-1}$) & (km s$^{-1}$) & \\
\hline

627 & & B7V & 403v & 1.1$\times 10^{-3}$ & $<$0.16 & & & Nearby star formation. J01\\

4881 && B9.5V & 168r & 1.9$\times 10^{-3}$ & {\bf 0.12$\pm$0.05} & -15.3 & -13.7$\pm$2 & S(0.98). Poss. cirrus. K02\\

6028 && A3V & 91v & 1.1$\times 10^{-3}$ & $<$0.15 & & & \\

9672 & 49 Cet & A1V & 61 & 8.7$\times 10^{-4}$ & {\bf 0.34$\pm$0.07} & 10.5 & 9.9 & D. Z95,S96,MS00\\

17081 & & B7IV & 135 & 1.3$\times 10^{-4}$ & $<$0.064 & & & \\

23362 && K5III & 308 & 7.9$\times 10^{-4}$ & $<$0.19 &&& Poss. cirrus. S96,K02\\

23680 && G5IV & 180 & 3.0$\times 10^{-3}$ & $<$0.15 &&& S96\\

31293 & AB Aur & A0V & 144 & 0.48 & {\bf 7.4$\pm$0.17} & 15.1 & 8$\pm$5 & D. MS97,M01\\

31648 & MWC480 & A5V & 131 & 0.2 & {\bf 2.74$\pm$0.06} & 14.5 & - & D. MS97,S01\\

32509 && A2V & 150 & 0.016 & $<$0.11 &&&\\

34282 && A0.5V & 400r & 0.39 & {\bf 1.1$\pm$0.12} & 16.2 & - & D. PDK03,S96\\

34700 && G0IV & 125r & 0.15 & {\bf 0.42$\pm$0.06} & 21.0 & - & D. S96,CWD98,SDB01,T04\\

35187 && A2V & 150 & 0.16 & $<$0.26 &&& Binary. J01,DC98.\\

35929 & & F0III & 400r & 0.011 & $<$0.048 &&& vdA98\\

36112 & MWC758 & A3V & 204 & 0.17 & {\bf 0.81$\pm$0.1} & 17.0 & 17.6$\pm$0.2 & S(2.4). MS97,T01,B99\\

36910 & CQ Tau & F5IV & 99 & 0.25 & {\bf 0.31$\pm$0.05} & 17.7 & 20 & D? N01,MS00\\

37806 & & B9V & 280v & 0.32 & $<$0.14 & && M98\\

38120 && B9V & 420r & 0.47 & {\bf 0.66$\pm$0.12} & 31.6 & - & S(1.4). CWD98\\

48682 && G0V & 16 & 1.2$\times 10^{-4}$ & $<$0.11 & & & \\

50138 & & B8V$^{(6)}$ & 289 & 0.6 & $<$0.074 & & & Poss. Be star\\

53833 && A9V & 267v & 0.02 & $<$0.25 & && IR not near star. SM00\\

58647 & & B9IV & 280 & 0.15 & $<$0.11 & && Poss. Be star\\

81515 && A5V & 106 & 6.1$\times 10^{-3}$ & $<$0.09 && & \\

98800 && K4V & 46 & 0.084 & $<0.09$ & && S96\\

102647 & $\beta$ Leo & A3V & 11 & $1.9 \times 10^{-5}$ & $<$0.18 & && J=2-1 CO. J01, D95\\

109085 && F2V & 18 & 6.9$\times 10^{-5}$ & $<$0.08 && & \\

121847 & 47 Hya & B8V & 104 & $2.4 \times 10^{-4}$  & $<$0.07 & &&\\

123247 & & B9.5V & 101 & 1.3$\times 10^{-4}$ & $<$0.3 & && \\

123356 & & G1V & 41 & 6.1$\times 10^{-3}$ & $<$0.07 & &&\\

135344 & SAO206462 & F4V & 84r & 0.44& {\bf 0.97$\pm$0.04} & 2.9 & -3$\pm$3 & D. S96,CWD98,T01,M01,D03\\

139365 & & B2.5V & 136 & 1.4$\times 10^{-5}$ & $<$0.063 & & &\\

139450 & & G0/1V & 73 & 2.4$\times 10^{-3}$ & - &&& Ambient emission\\

139614 & & A7V & 157r & 0.39 & {\bf 0.47$\pm$0.11} & 3.3 & 3$\pm$1& S? S96,M01,D03\\

141569 & & A0V & 99 &  8.4$\times 10^{-3}$ & {\bf 0.76$\pm$0.055} & -7.5 & -6$\pm$5 & D. S96, L03\\

142114 & & B2.5V & 132 &  7.5$\times 10^{-5}$ & $<$0.07 & & &\\

142165 & & B5V & 127 &  4.0$\times 10^{-5}$ & $<$0.064 & & &\\

142666 & & A8V & 116r & 0.28 & {\bf 0.72$\pm$0.14} & -5.0 & 3$\pm$1 & S?(4.9). S96,M01,D03\\

143006 & & G5V & 82r & 0.37 & {\bf 0.15$\pm$0.06} & -0.8 & -0.9$\pm$0.3 & S(0.95). J01,S96\\

143018 & & B1V & 140 &  1.5$\times 10^{-3}$ & $<$0.14 & & & \\

144432 & & A9V & 200 & 0.26 & {\bf 0.31$\pm$0.13} & -2.2 & 2$\pm$3 & D? S96,M01,D03 \\

145718 & V718 Sco & A8IV$^{(6)}$ & 130 & 0.1 & {\bf 0.45$\pm$0.23} & -4.2 & - & D? Marginal det. Z95,T01\\

150193 & MWC863 & A1V & 150 & 0.15 & $<$0.07 & && Poss. ambient emission. M01 \\

155826 & & G0V & 30 &  6.5$\times 10^{-4}$ & - & && Bright ambient em. L02\\

163296 & & A1V & 122 & 0.16 & {\bf 4.3$\pm$0.1} & -6.4 & -4$\pm$1 & D. MS97,M01,J01 \\

169142 & & A5V & 145r & 0.1 & {\bf 1.7$\pm$0.13} & -3.1 & -3$\pm$2 & S(1.6). M01,D03\\

179218 & MWC614 & A0IV & 243 & 0.62 & {\bf 0.6$\pm$0.12} & 16.1 & -3$\pm$5 & D. M01\\

190073 & MWC325 & A2IV & 400 & 0.4$^{(5)}$ & $<$0.08 & && \\

191089 & & F5V & 53 &  2.3$\times 10^{-3}$ & $<$0.25 & &&\\

212676 & & B9V & 670 &  4.7$\times 10^{-3}$ & $<$0.08 & && CWD98\\

214953 & & G0V & 23 &  2.3$\times 10^{-4}$ & $<$0.28 & &&\\

215592 & & A0V$^{(6)}$ & 590r & 0.048 & {\bf 0.18$\pm$0.12} & 1.7 & 2.7$\pm$5 & Marginal detection. CWD98\\

224648 & & B9V$^{(6)}$ & 300 & $8.7 \times 10^{-3}$ & $<$0.085 & && \\

232344 & & B9V$^{(6)}$ & 450 & 0.086 & $<$0.09 & &&\\

233517 & & K5III & 600r & 0.057 & {\bf 0.21$\pm$0.11} & 36.0 & 46.5$\pm$1 & S(3.6). S01,F96\\

287841 & V346 Ori & A2IV & 400 & 0.1 & $<$0.06 & && \\

293782 & UX Ori & A4IV & 430 & 0.35 & {\bf 0.21$\pm$0.07} & 17.2 & 18.3 & S(0.98). CO at v=27. N01\\

344361 & WW Vul & A2IV & 370 & 0.42 & $<$0.095 & && N01\\

- & VX Cas & A0V & 760 & 0.3$^{(5)}$ & $<$0.09 & && \\

- & TW Hya & K7e & 56 & 0.3 & {\bf 2.16$\pm$0.2} & 12.35 & 12.6 & S. K97 \\

\\

\hline
\end{tabular}
\end{center}
\end{table*}

\newpage

Notes to Table 1: \\

(1) Spectral types are either from Moro et al. (2001), from the references
given, or from the Simbad database.\\
(2) Distances are normally from Hipparcos, except those marked `r', which are
taken from the given reference, or `v' estimated from visual magnitude,
uncorrected for reddening.\\
(3) Fraction excess is taken from the literature, or where not available,
extimated from a fit to the SED (see text for details).\\
(4) Intensity units are main beam brightness temperature (using
$\eta_{\mathrm{mb}}=0.62$).
To calculate the upper limits, the rms noise has been measured with
10 km s$^{-1}$ channels; 1-$\sigma$ values are given.\\
(5) Estimate of fractional excess is uncertain, due to wide range of
temperatures in excess emission, possibly high extinction to star and/or
incomplete wavelength coverage.\\
(6) Luminosity classification based on distance and optical magnitude.\\
(7) For targets with detected CO, an indication is given of whether the line
is double-peaked (D) or single-peaked (S). In the case of single lines, the
linewidth (FWHM) is also given (in km s$^{-1}$).\\

References:

B99: Beskrovnaya et al., (1999);
CWD98: Coulson et al. (1998);
D03: Dominik et al. (2003);
DC98: Dunkin \& Crawford (1998);
F96: Fekel et al. (1996);
G00: Greaves et al. (2000);
J01: Jayawardhana et al. (2001);
K02: Kalas et al. (2002);
K97: Kastner et al. (1997);
L02: Lisse et al., (2002);
L03: Li \& Lunine (2003);
M98: Malfait et al., (1998);
M01: Meeus et al. (2001);
MS97: Mannings \& Sargent (1997);
MS00: Mannings \& Sargent (2000);
N01: Natta et al. (2001);
PDK03: Pi\'etu, Dutrey, Kahane (2003);
S01: Simon et al. (2001);
SDB01: Sylvester, Dunkin \& Barlow (2001);
S96: Sylvester et al. (1996);
SM00: Sylvester \& Mannings (2000);
T01: Thi et al. (2001);
T04: Torres (2004);
vdA98: van den Ancker et al. (1998);
Z95: Zuckerman et al. (1995).

\vskip 5mm

\section{Detectability of CO}

A total of 59 targets were observed, of which 21 were detected in CO
with 2 additional marginal detections (at the 2-$\sigma$ level).
In Fig.~1, we plot $I_{\mathrm{CO}}$, the measured CO intensity or 1-$\sigma$ limit
normalised to a distance of 100~pc,
as a function of the fractional infrared excess $f$.
The normalisation of the CO line intensity
assumes the emission to be smaller than the beam; the emitting
regions are typically a few hundred au in radius (see below), which compares
with the beam size of 1400 au at a distance of 100 pc.
As the $^{12}$CO line is likely to be optically thick (see below),
the normalised CO intensity can be used to estimate a lower limit to
the gas mass.
Assuming optically thin emission and that the gas
is in LTE, the total gas mass would be given by (e.g. Thi et al., 2001):

\begin{equation}
M_g \approx 10^{-4} (T_{ex}+0.89)/e^{16.02/T_{ex}} I_{\mathrm{CO}}
\end{equation}

where $T_{ex}$ is the excitation temperature, $M_g$ the mass
in units of Jupiter masses (M$_J$), the $^{12}$CO/H$_2$ abundance
ratio is assumed to be $5 \times 10^{-5}$ and $I_{CO}$ is normalised to 100~pc
(see above). Most emission arises from the outer regions of these objects, 
where the dust kinetic temperature, $T_{dust}$,
is between $\sim$30K (e.g. Beckwith \& Sargent, 1993) and 
$\sim$100K (e.g. Dent et al., 2000). Assuming H$_2$ densities greater
than 10$^3 \mathrm{cm}^{-3}$, then the gas and dust are sufficiently coupled for
$T_{dust} \approx T_{ex}$. From eqn. (1), the likely range of
temperatures will affect the
derived mass limit by no more than a factor of 2
and so, adopting an excitation temperature of 60K, the horizontal
line at $I_{CO} = 0.1$ on Fig.~1 corresponds to a minimum mass of
$10^{-3} \mathrm{M}_J$.

Also included in Fig.~1 are published limits to the
J=2-1 CO emission from the Vega-excess stars Vega,
$\beta$ Pic, Fomalhaut and HR4796 (Dent et al., 1995; Liseau, 1999);
we have assumed the same brightness
temperature limits in both the J=3-2 and J=2-1 transitions.

The plot shows a clear link between the detectability of CO and a high
fractional excess. We find 18 of the 27 objects with $f \geq 0.1$ are
detected (upper right of the figure). Including the upper limits,
the results indicate that 67 - 100 per cent of objects with
$f > 0.1$ and 56 - 98 per cent of objects with $f > 0.01$ have
significant CO emission (defined as where
$I_{CO}.(D/100\mathrm{pc})^2 \geq 0.1 \mathrm{K km s}^{-1}$).
The remaining stars with $f < 0.01$ are generally not detected in CO (only
2 or possibly 3 clear detections out of 27 objects with
uncontaminated spectra). Ignoring the three very distant objects with
poor scaled upper limits, this implies that
$\sim$8 - 20 per cent of low-excess objects ($f < 0.01$) have
$I_{CO}.(D/100 \mathrm{pc})^2$$ \geq $$ 0.2 \mathrm{K km s}^{-1}$.
However there are some notable exceptions to this general rule
which are worth looking at individually.

HD98800 is the only star with $f \geq 0.01$ where the CO limit is
significantly lower (ie $\geq 3\sigma$) than all the others.
As will be shown later, this indicates a relatively small limit
to the surface area of circumstellar gas, although it
may be linked to the fact that this is a quadruple system.
Conversely, there are two (possibly three) discs with low
$f$ ($<$0.01) which have relatively bright CO emission:
HD141569, which has a clear double-peaked line and a scattered light
disc, HD9672 (also double-peaked)
and HD4881. These will be discussed individually in later sections.

\epsfverbosetrue
\epsfxsize=14.0 cm
\begin{figure*}
\center{
\leavevmode
\epsfbox{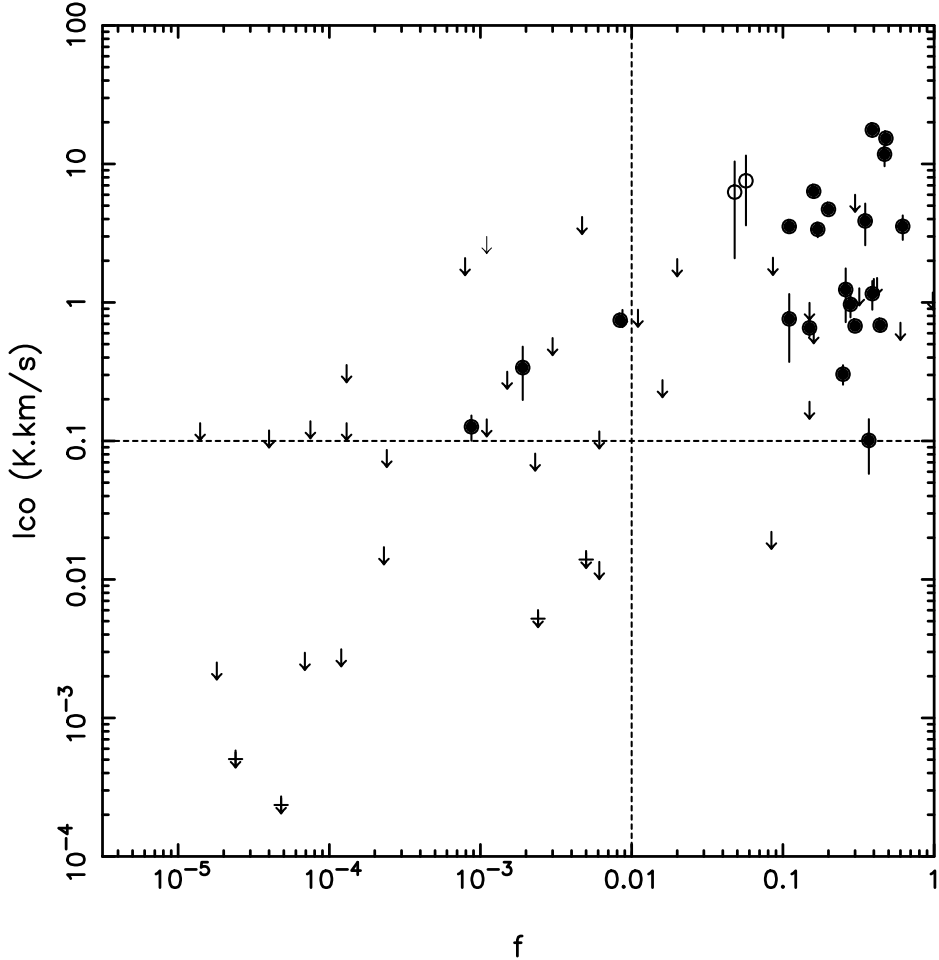}
\caption { 
Integrated CO intensity for target stars
(T$_{\mathrm{mb}}$dv, in units of K km s$^{-1}$) normalised to a
distance of 100 pc, plotted against the fractional excess of continuum
emission above that of the stellar photosphere. Objects with clear
CO detections are depicted as filled circles and those with uncertain
identifications
as open circles; the 1-$\sigma$ upper limits of the remainder are also shown.
Also included are published CO upper limits of nearby
archetypical Vega excess stars (Dent et al., 1995; Liseau, 1999);
these are shown as upper limits with a cross. The horizontal
dashed line represents a minimum gas mass of $10^{-3} \mathrm{M}_J$,
and vertical
line represents the critical excess for CO emission (see text for details).}
}
\end{figure*}

For a reprocessing disc viewed near face-on, the infrared excess is
approximately the fraction of stellar radiation intercepted by the disc.
To first order this is the effective solid angle subtended by the
disc as seen from the star so, assuming a disc of constant opening angle
$\theta$, and mean optical depth through the disc to optical
stellar photons of $\tau_V$, the excess $f$ is given by (e.g. Backman \&
Paresce, 1993):

\begin{equation}
f = (1-e^{-\tau_V}) sin ( \theta/2 )
\end{equation}

In a typical photodissociation region, CO can survive at $A_V\geq$2-6
(Hollenbach et al., 1991).
As $\tau_V \sim A_V$, then assuming that photons from the central star are the
dominant cause of CO dissociation (van Zadelhoff et al., 2003),
we can estimate the disc opening angle required for significant
CO to be present. The results of this survey imply that $>$67 per cent of discs of
opening angle $\geq 12\degree$ ($f \geq 0.1$) will have gas masses
$\geq 10^{-3} \mathrm{M}_J$. Including the upper limits and ignoring the
quadruple system HD98800, the results are consistent with
{\em all} discs of $f \geq 0.01$, or $\theta \geq 1\degree$, having
at least $10^{-3} \mathrm{M}_J$ of molecular gas.

The low extinction along the line of sight
towards the stars with detected circumstellar CO (typically $A_V \sim 1$ mag;
van den Ancker et al., 1998) means we are unlikely to be looking through the
disc itself, so the inclination $i < (90-\theta/2)$. Assuming the sample of
discs is randomly-oriented on the sky, the mean inclination would
be $60\degree$, giving a mean opening angle $\theta$ of $< 30\degree$.
For realistic discs, the photons may be intercepted at an inner hot
bulge (e.g. Natta et al., 2001) rather than an outer flaring region, although
either structure would result in an increase in the total excess
$f$, as well as shielding more distant gas in the disc from
UV photodissociation; as a result the above relation would still apply.

The gas discs with the lowest $f$ are HD141569, HD4881 and HD9672;
from eqn. 1, their opening angles range from $\sim 1\degree$ down to as small
as $0.1\degree$.
There is no evidence for strong near-infrared excesses in these stars
(e.g. Sylvester et al., 1986) as predicted by the hot inner bulge models,
which would suggest the optically thick regions of
their discs have relatively large radii as well as being physically thin.
Such structures are even thinner than the opening angle of 7$\degree$ seen
in the edge-on $\beta$ Pictoris dust disc (Heap et al., 2000), and
the detections of CO imply it can also exist in such thin layers.

\section{Spectral profiles}

Emission from CO around five of the brightest isolated
HAeBe stars have been resolved with interferometers
(Mannings \& Sargent, 1998, 2000; Pi\'etu et al., 2003; Augereau
et al., 2004). In all cases
the material appears to be located in Keplerian discs
with outer radii of a few hundred au; however the angular
sizes are often only marginally larger than
the existing interferometer beam sizes.

The CO emission intensity from most of the objects in the current survey
is $\leq 1 \mathrm{K km s}^{-1}$,
and is difficult to image with existing inteferometers. Consequently
we have to rely on interpretation of the emission line profiles
to derive kinematic and geometric information about the discs. 
It was noted by Beckwith \& Sargent (1993) and others
that low-J $^{12}$CO lines in discs around young stars
are likely to be optically thick, and so effectively trace the
gas temperature on the disc surface rather than the total mass.
A double-peaked line profile is a characteristic of such emission.
Of the 21 stars with CO emission, 9 and possibly as many as 12
have clear double-peaked line profiles (Table~1).
Formally a double-peaked profile can also arise from self-absorption
in an extended cool outer envelopes (as found in obscured young stellar
objects). However all our target stars have low optical
extinction, suggesting that this effect is unlikely to be important.

A small number of objects (HD4881, HD143006 and UX Ori) have narrow lines
($\leq 1 $km s$^{-1}$), consistent with emission from an ambient cloud.
However, even in these cases the CO and stellar velocities are similar,
suggesting the gas may be closely associated with the star.
Four stars have single-peaked lines significantly broader than
typical ambient gas emission; this is thought to be from a disc
viewed close to edge-on (see below). The remaining lines have insufficient
s:n to classify their shapes, although most are relatively broad.
In the modelling described below, we have assumed all our detections
are due to disc emission, and used the line profiles to
estimate system parameters.

\subsection{Modelling program}

As recognised by, for example, Omodaka et al. (1992) and Beckwith \& Sargent
(1993), it is possible to derive some basic parameters of
a disc, such as size and inclination, from fitting
just the spectral profile of
an optically thick tracer. The high mean density means the CO line
is thermalised, allowing a simple model to be applied which assumes that the
local excitation temperature is unaffected by emission from other parts
of the disc. Significantly more complicated models of line emission
from discs have been developed more recently which include
the effects of chemistry, radiative transfer, different dust characteristics,
more complex disc structures, and UV photodissociation (e.g. van Zadelhoff et
al., 2001; Aikawa et al., 2002).
However, each of these introduces many new uncertain parameters,
and so in the absence of additional data for most targets, we have
applied just a basic model to the J=3-2 CO profiles, minimising
the $\chi^2$ fit to the data.

The present model assumes the disc is in Keplerian rotation viewed at
inclination $i$ ($i$=0 implies a face-on geometry), with a 
an effective opening angle, $\theta$, at radius 100~au.
Volume density in the mid-plane falls as $r^{-3}$, and the disc is flared, so
the (Gaussian) scale height, $z_o$, increases as $r^{1.5}$ (see
Beckwith \& Sargent, 1993).
However, the CO profile is not sensitive to either of these power laws.
The gas density $n_{\mathrm{H}_2}$ at radius $r$ and height $z$ above
the disc plane, between $R_{in}$ and $R_{out}$ is then given by:

\begin{equation}
n_{\mathrm{H}_2} = n_{0} (r/r_{0})^{-3} e^{-(z/z_{0})^2}
\end{equation}

where the scale height is expressed as:

\begin{equation}
($${z_{0}}\over{100 \mathrm{au}}$$) = \mathrm{tan}(\theta) $$({{r}\over{100 \mathrm{au}}})^{1.5}$$
\end{equation}

The density is assumed to be zero for $r < R_{in}$ and $r > R_{out}$.
and $n_{0}$ and $r_{0}$ are scaling factors for the total disc mass,
obtained from the dust mass
assuming a constant gas:dust ratio of 100. Dust masses are obtained
from the literature, mostly from fits to the continuum SED.
The narrow features in several of the spectra suggests that
the effect of turbulent line broadening is negligible; consequently
only thermal broadening is included (see also Beckwith \& Sargent;
Pi\'etu et al., 2003).
The model assumes a constant gas:dust ratio without the effects of
chemistry; Aikawa \& Herbst (2001) indicate that chemical reactions
in the disc do not significantly alter typical emergent lines profiles
of molecules such as $^{12}$CO, as the emission is so optically
thick. However, the CO abundance
can be affected by the local UV photoionisation rate, which depends on
distance and extinction to the star and the stellar spectral
type, as well as the interstellar radiation field and optical depth to
the surface of the disc. The stellar photodissociation is based on
an analytical fit to the model of Hollenbach et al. (1991).
Interstellar photodissociation is approximated
by effectively removing emission from the disc surface down to A$_V = 3$ mag.
However, in most cases the density is sufficiently high that the UV has
little effect on the CO abundance throughout most of the disc volume.

As the mean densities for most discs under investigation are
typically $>>10^4 \mathrm{cm}^{-3}$, then
the gas and dust will be in thermal equilibrium; the dust temperature
is assumed to be that of small grains radiatively heated
the star (e.g. Aikawa \& Herbst, 2001). The excitation temperature is then given
by:

\begin{equation}
T_{ex} = 282 . $$({L \over{\Lsun}})$$^{0.2}.$$({d \over{\mathrm{au}}})$$^{-0.4}.$$({a \over{\mu \mathrm{m}}})$$^{-0.2}
\end{equation}

Smaller grains, such as classical
ISM dust, will be hotter, but offsetting this is the possibility that
the outer disc may be shielded from the star by
the inner regions and so would be cooler.
The extinction to the star depends on the flaring angle and detailed
structure of the inner disc; for a flared disc for example,
the dust higher above the disc plane will be hotter.
The effective CO excitation temperature at a particular radius
is approximately that at the `$\tau$=1' surface (van Zadelhoff et al.,
2001). The height of this surface above the disc plane
depends on $\theta$ and the inclination angle $i$ as
well as the disc structure.
In the case of a face-on disc, comparison of the temperatures at
the $\tau$=1 surface (see van Zadelhoff et al., 2001, their Figure~6)
with the temperatures assumed in our present
model indicates that we may be underestimating the temperature at
a given radius by a factor of between 1.0 and 2.0.
For more inclined discs, however, 
we will see deeper into the gas, and so the temperature at the
`$\tau$=1' surface will actually be lower. This suggests that the
uncertainty in $T_{ex}$ is less
than a factor of 2; for unresolved discs in an optically thick line,
this implies the disc radius
would then be in error by a factor of $\leq 1.5$, which gives some
indication of the possible uncertainty of $R_{out}$.

\subsection{Line profile fitting}

The spectra of the targets with significant
detections of CO are shown in Figure~2, along
with the best fit models. Table~2 summarises the modelling results
for these objects; also given is the derived disc angular diameter.
In some cases the parameters are not well constrained
because of the low s:n of the data - this is indicated in the table. 
Dust masses marked (1) are scaled to the adopted distances. Lines with
a clear double-peaked profile are identified by `D'.

\epsfverbosetrue
\epsfxsize=16.0 cm
\begin{figure*}
\center{
\leavevmode
\epsfbox{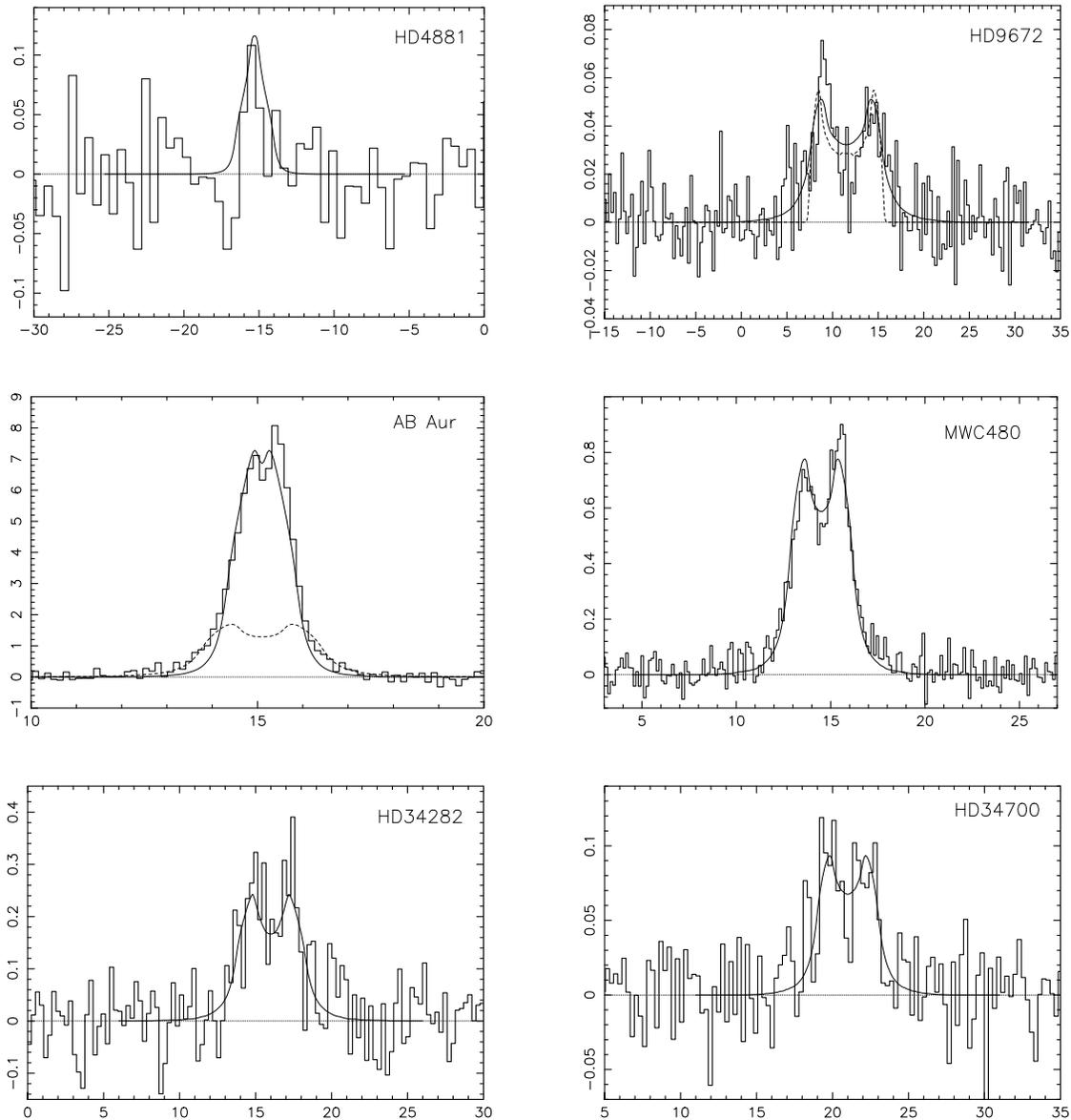}
\caption { 
(a-d) Spectra of J=3-2 $^{12}$CO for the objects where emission was detected
(histogram). Also shown are the best fit models (solid line); in some
cases an alternative model is shown by the dashed lines - see discussion
of individual objects for details.
Model parameters are given in Table 2 and described in more detail
in the text for individual objects. Velocity scale is km s$^{-1}$
(Heliocentric reference), and
intensity unit is main beam brightness temperature (K); note that the ranges
on the axes vary from source to source.}
}
\end{figure*}

\epsfverbosetrue
\epsfxsize=16.0 cm
\addtocounter{figure}{-1}
\begin{figure*}
\center{
\leavevmode
\epsfbox{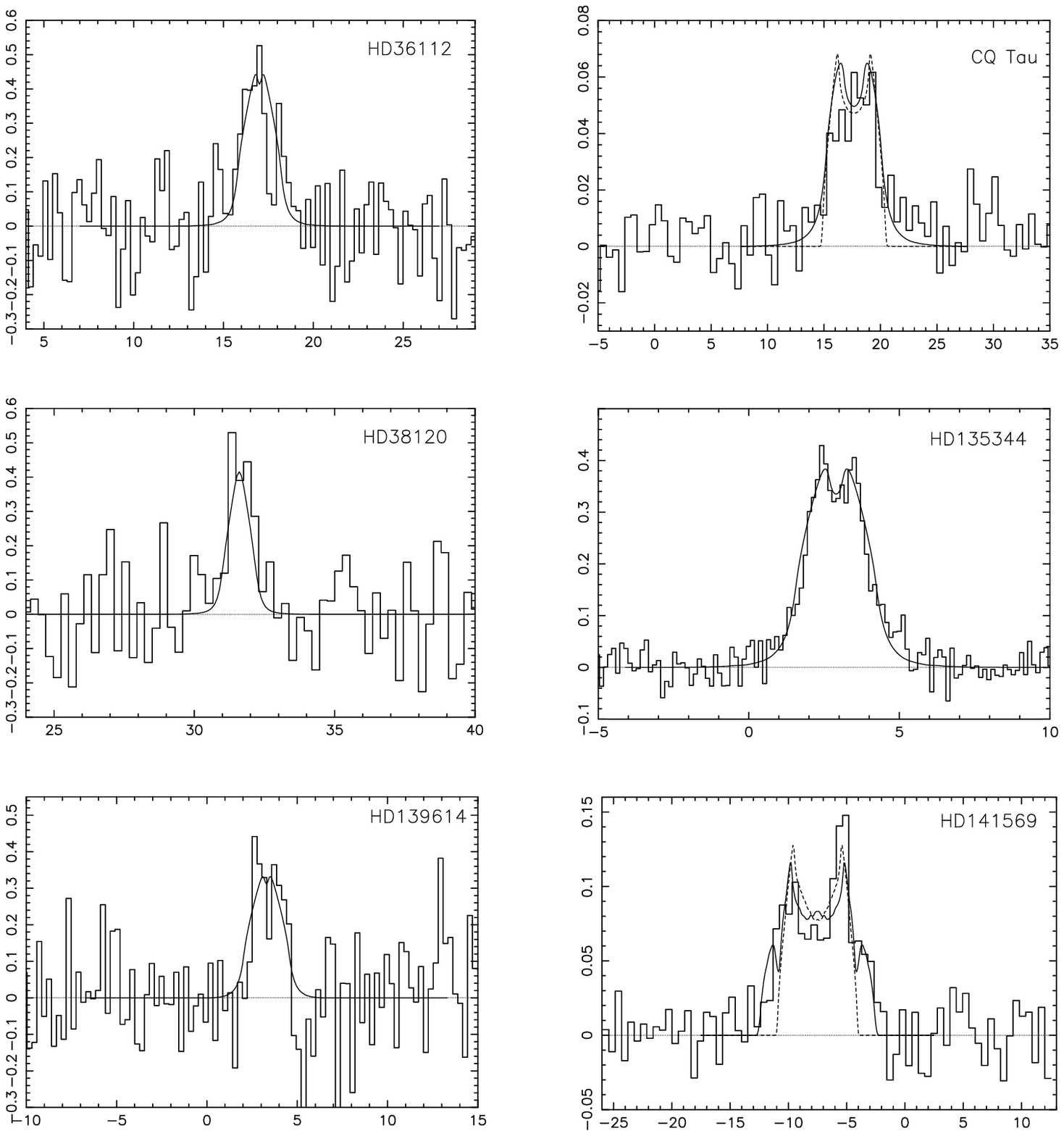}
\caption { (b) }
}
\end{figure*}

\epsfverbosetrue
\epsfxsize=16.0 cm
\addtocounter{figure}{-1}
\begin{figure*}
\center{
\leavevmode
\epsfbox{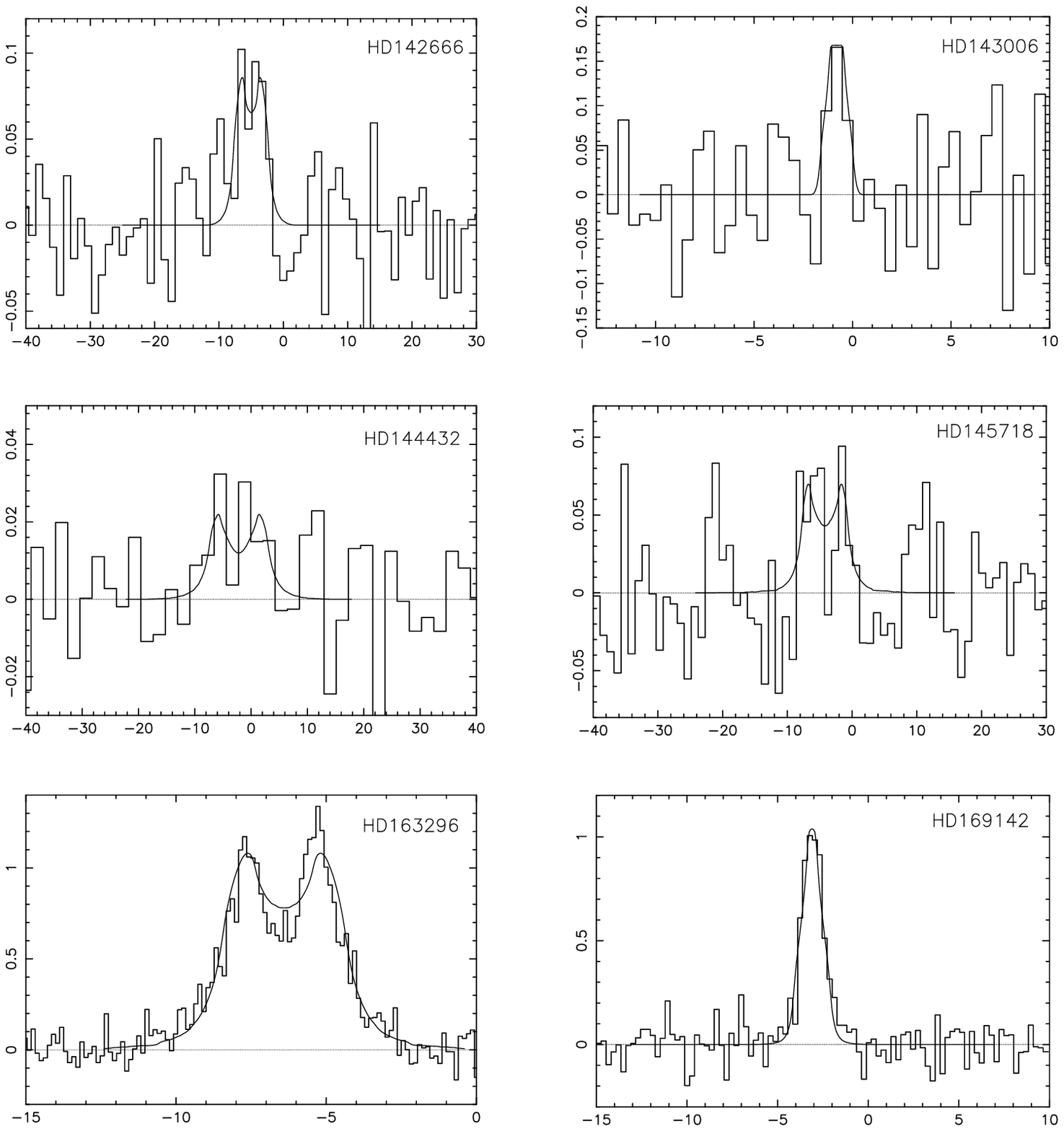}
\caption { (c) }
}
\end{figure*}

\epsfverbosetrue
\epsfxsize=16.0 cm
\addtocounter{figure}{-1}
\begin{figure*}
\center{
\leavevmode
\epsfbox{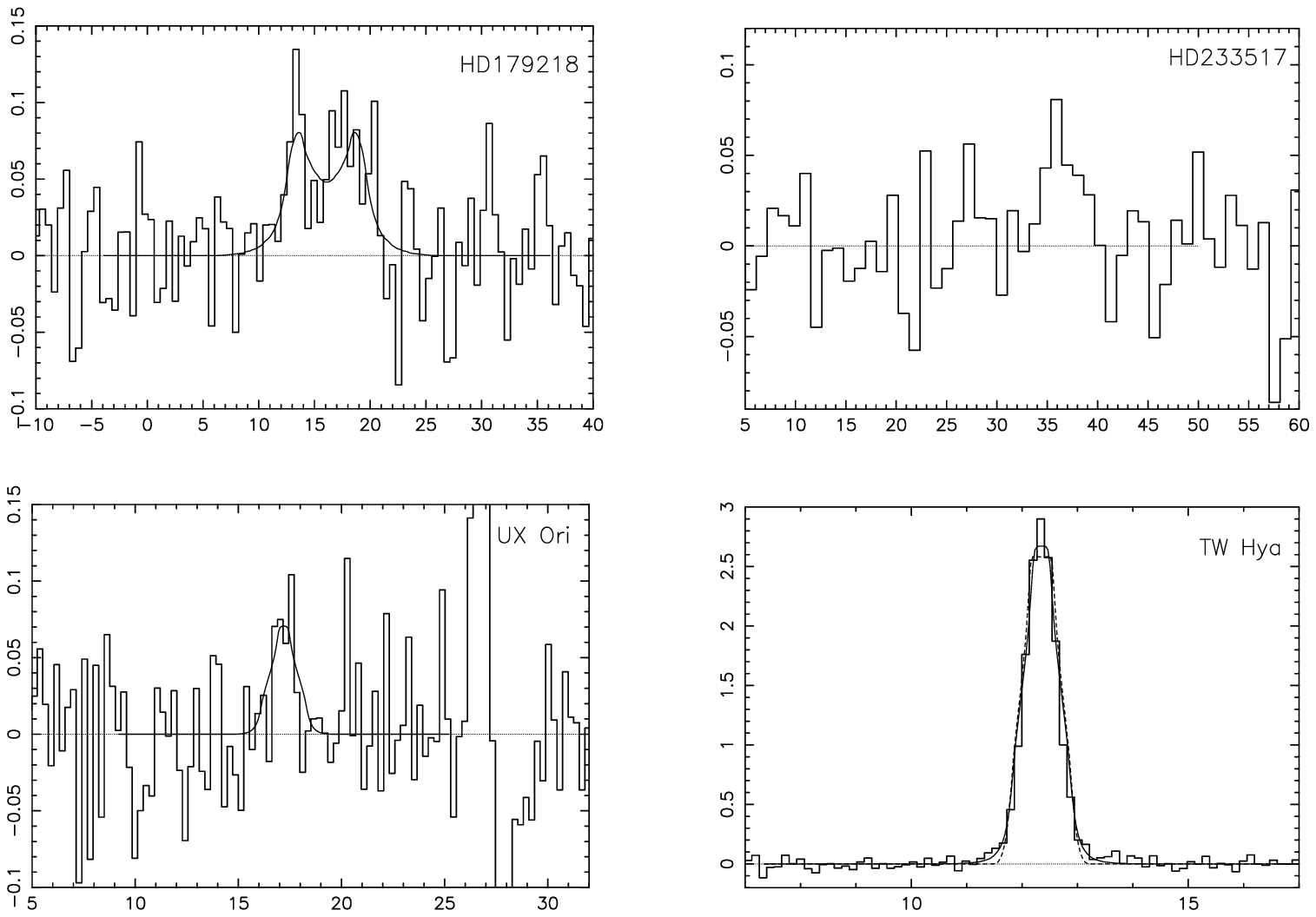}
\caption { (d) }
}
\end{figure*}

In most cases, the CO profiles can be fitted reasonably well
by the simple disc model by a suitable choice of $i$ and $R_{out}$.
As the line
emission is generally optically thick from all regions of the disc, the
disc outer radius determines the intensity to first order, and
the disc inclination mainly affects the overall line width.
The inner radius is constrained by the line
wings although, apart from the closest objects, beam dilution of emission
from this compact region means $R_{in}$ is not well constrained.
The opening angle $\theta$ has a small effect on the line intensity and
shape at the lower relative velocities, although in most cases we cannot
constrain it, and have simply set it to a fixed value.

The spectra from nine objects
listed in Table~2 have clear double-peaked spectra consistent with a disc.
Another nine objects have single-peaked lines, although only
a small number of these (HD4881, HD143006 and UX Ori)
have narrow line widths more consistent with ambient cloud
emission ($<$ 1 km s$^{-1}$). The remaining have more ambiguous line profiles
(CQ Tau, HD139614, HD144432, HD145718), although all are relatively
broad suggesting they arise from a disc rather than an ambient cloud.
Apart from HD179218 and HD233517 the stellar velocities, if known,
agree with those of the CO, confirming
the gas is associated with the stars. Assuming all the
single-peaked profiles arise from discs viewed almost face-on, then the average
outer radii of these objects is 110~au, compared with 170~au for
the double-peaked and broad spectra; the similarity of these values also
suggests a common origin for all emission.
In the following sections, we discuss the results for individual objects.

\begin{table*}
\begin{center}
\caption{\sf Stars with CO detections: best fit model parameters.}
\begin{tabular}[t]{clllllll} \hline\
Star & Dust mass &  $R_{in}$ & $R_{out}$ & $D_{out}$ & $i$ & $\theta$ & Notes$^{(2)}$ \\
   & ($\Msun$) & (au)     & (au)     & (arcsec) &     &          &        \\
\hline

HD4881 & $3.9 \times 10^{-5}$ & - & 35 &0.4& $\leq$5 & $\leq$10 & \\

HD9672 & $4 \times 10^{-6}$ & $<$5 & 17$\pm 5$ &0.6& 16$\pm$3 & 10 & D\\

AB Aur & $2.1 \times 10^{-4}$ & $\leq$50 & 600$\pm$50 &8.4& 12$\pm$2 & 5 & D; Poss. 300au inner disc - see text\\

MWC480 & $2.0 \times 10^{-4}$ & $\leq$20 & 245$\pm$30 &3.8& 28$\pm$2 & 10 & D\\

HD34282 & $1.1 \times 10^{-3}$ & $\leq$50 & 380$\pm$20 &1.8& 50$\pm$5 & 10 & D\\

HD34700 & $2.1 \times 10^{-5}$$^{(1)}$ & $\leq$30 & 80$\pm$10 &1.3& 25$\pm$2 & 10 & D\\

HD36112  & $2.9 \times 10^{-4}$ & - & 170$\pm$30 &1.7& $\leq$10 & 10 & Single broad line\\

CQ Tau  & $2.2 \times 10^{-4}$ & $\leq$10 & 30$\pm$5 &0.7& 14 & 10 & Broad line\\

HD38120  & $10^{-5}$ & - & 300$\pm$50 &1.2& $<$8 & 10 & \\

HD135344  & $10^{-4}$ & $\leq$10 & 75$\pm$5 &1.8& 11$\pm$2 & 10 & D\\

HD139614  & $10^{-4}$ & $\leq$20 & 110$\pm$3 &1.4& $\leq$10 & 10 &\\

HD141569  & $10^{-5}$ & 90$\pm$5 & 250$\pm$2 &5.0& 52 & 0.5 & D; Double ring structure - see text\\

HD142666  & $3 \times 10^{-4}$ & - & 45$\pm$10 &0.8& 18$\pm$5 & 10 & Single broad line\\

HD143006  & $1.6 \times 10^{-5}$ & - & 35$\pm$5 &0.8& $\leq$3 & 10 &\\

HD144432  & $2 \times 10^{-3}$ & - & 60$\pm$20 &0.6& 48$\pm$10 & 10 & Low s:n; broad line? \\

HD145718  & $4.5 \times 10^{-5}$ & - & 60$\pm$30 &0.9& 32$\pm$10 & 10 & Low s:n \\

HD163296  & $5 \times 10^{-4}$ & $\leq$20 & 245$\pm$20 &4.0& 30$\pm$2 & $\leq$10 & D\\

HD169142  & $10^{-3}$ & - & 130$\pm$10 &1.8& $\leq$5 & 10 & \\

HD179218  & $10^{-4}$ & - & 120$\pm$20 &1.9& 40$\pm$10 & 10 & D; Stellar and CO velocity differ\\

HD233517  & $1.7 \times 10^{-4}$$^{(1)}$  & - & - & - & - & - & Stellar and CO velocity differ\\

UX Ori  & $4.2 \times 10^{-4}$ & - & 95 & 0.4 & $\leq$8 & 10 & Additional ambient gas component\\

TW Hya  & $3.0 \times 10^{-4}$ & $\leq$30 & 160$\pm$10 & 5.7 & 5$\pm$2 & $\leq$10 & \\

\\
 
\hline
\end{tabular}
\end{center}
\end{table*}

\subsubsection{HD9672}

HD9672 (49 Cet) is one of only three debris discs with
$f > 10^{-3}$ (e.g. Jayawardhana et al., 2001), the other two
being HR4796 and $\beta$ Pictoris.
A weak CO line was detected by Zuckerman et al. (1995), making this
not only the only bona-fide debris disc with detectable CO, but
also one of the closest gas-rich discs (only TW~Hya is closer). Fig.~2a
confirms the emission centroid lies close to
the stellar velocity (9.86 km s$^{-1}$)
and shows the line profile to be double-peaked.

Weinberger et al. (1999) saw no evidence of scattered light at radii
beyond 1.6 arcsec ($\sim$100~au), although
Jayawardhana et al. (2001) did detect mid-infrared emission extending
to a radius of $\sim$ 50 au.
The relatively weak CO line brightness implies the emission region
likely has a small surface area. The data are consistent with a
compact disc (outer radius $\sim$17~au) inclined at $16\degree$.
The relatively high velocity wings imply gas is present at radii
$\leq$5~au.
A more inclined ring ($i \sim 35\degree$) of radius 50~au
(equal to that seen in the mid-infrared) would
produce a similar separation of the line peaks but
does not reproduce the higher velocity wings
(see dashed line on the spectrum in Fig.~2a).

\subsubsection{AB Aur}

A disc was resolved around this star using mm-wave interferometry
in the J=1-0 $^{13}$CO line (Mannings \& Sargent, 1997). The
beam size was 4-5 arcsec ($\sim$720 au) and
the derived outer radius was 450~au.
HST images show a scattering disc extending to a radius of
$\sim 8$ arcsec (Grady et al., 1999), and an extended
arc of dust at $r \sim 10-20$ arcsec, likely due to remnant ambient gas.
Grady et al. suggest that the disc is almost face-on ($i < 45\degree$),
and may be flared, judging by the radial emission profile.

Our fit using an outer radius of 600~au (Fig.~2a) shows reasonable
agreement with the spectrum at low relative velocities. 
Notably, the narrow linewidth implies that the disc must be almost face-on
($i \sim 12\degree$), agreeing with the
symmetrical optical image, rather than the higher inclination derived
from the interferometer map.
But this simple model cannot explain the line wings seen out to
relative velocities of $\pm 2$km s$^{-1}$. This implies that either the
inclination of the inner region of the disc is higher (a model with
$i = 17\degree$, $i\sim300$~au can fit the profile of the wings alone, and
is shown by the dashed line in Fig.~2a), the
low-velocity emission is dominated by ambient gas, or there is an
additional higher-velocity component such as a molecular outflow. 
The presence of highly extended, possibly ambient material in the
scattered light images suggests that the narrow
central component in the $^{12}$CO line may not be from the disc itself;
consequently $i=17\degree$ and $R_{out}\sim 300$~au may represent the true
values of the circumstellar disc.

\subsubsection{MWC480}

This disc has been resolved in J=2-1 CO using interferometry with
a 1.8 arcsec beam (Mannings et al., 1997). By comparing with
a two-component Gaussian model of surface brightness, they derived an
outer cutoff of $\sim 650$ au (after scaling to the distance in Table~1).
However, our data (Fig.~2a) are consistent with a disc model with outer radius
$\sim$245 au; the reason for this discrepancy is thought to be
differences in the model and possibly the increased sensitivity
of their J=2-1 data to extended lower-density extended regions.

In addition to the double-peaked shape, two other features in
the deep CO spectrum are worthy of note. One is that emission is
detected out to relative velocities of $\pm 3.5$km s$^{-1}$, implying
the gas disc has an inner radius smaller than $20$~au. The
second is that the line profile
is clearly asymmetrical; a similar asymmetry was seen in the
reconstructed interferometric spectrum of J=2-1 $^{12}$CO in Mannings
et al. and the $^{13}$CO single-dish spectrum in Thi et al. (2004),
although the s:n of both of these spectra are lower than shown in Fig.~2a.
A possible explanation is that the asymmetry
is due to systematic telescope pointing offsets. With a 14 arcsec
Gaussian beam, the observed 30 per cent difference in intensities
of the two peaks in Fig. 2a
would require a consistent pointing offset of 5 arcsec
along the disc plane. Not only is this significantly larger than normal
telescope pointing errors, but also the data were taken from runs
on several different nights with consistent results.
We have instead attempted to fit this asymmetry
by imposing a sinusoidal azimuthal deviation in either the density or
temperature around the disc.  As the CO emission is optically
thick over most of the disc, implausibly large deviations in density
of $\geq 99$ per cent are necessary to explain the asymmetric spectra.
However, the peak-to-peak temperature difference required
to match the line profile is 30 per cent.
The origin of such a putative temperature difference could be 
non-axisymmetric variations in the disc structure; for example
one sector may be more flared and exposed to the stellar radiation.

This asymmetry needs further confirmation with high s:n
data in other molecular lines, along with more detailed modelling
including a fuller treatment of dust temperature and structure
throughout the disc.

\subsubsection{HD34282}

A detailed model has been derived from interferometric data by
Pi\'etu et al. (2003), including a revision to the distance and hence
the luminosity of this star (see Table~1).
The results from fitting our single-dish spectrum give
an inclination consistent with their data, although the outer radius
from our fitting is 360 au, a factor of $\sim$2 smaller than theirs.
The peak line brightness in Fig.~2a is 0.3K; if the disc is optically
thick with a mean brightness temperature of 30K, this would suggest a beam
filling factor of 1 per cent and hence a radius of
$\sim 300$au, more consistent with the current model.

\subsubsection{CQ Tau}

This object is thought to be a UX Ori star, suggesting the
disc is viewed close to edge-on, and an inclination of 66$\degree$ was
estimated based on a model of the variability by Natta \& Whitney (2000).
An upper limit to the gaseous disc radius of 85~au was measured using
interferometry (Mannings \& Sargent, 2000), although Testi et al. (2003)
resolved the dust continuum emission at 7mm, finding a radius of 100-200 au. 
The single-dish CO spectrum shows a relatively broad but weak line (Fig.~2b),
which can be fitted with a contiguous disc of outer radius 30~au (inner radius
$\leq$10~au) and
inclination 14$\degree$ (the solid line in the figure). But this inclination
is inconsistent with previous estimates based on the variability.
A larger, thin and somewhat more inclined ring model would fit the observed
brightness temperature and width of the line core, and be consistent with
the maximum dust radius ($R_{out} = 80$ au, $R_{in} = 70$ au,
$i = 30\degree$, $\theta=2\degree$; see dashed line in Fig.~2b).
However, there is weak evidence of broad
line wings, which are not reproduced by such a thin 
ring; furthermore a ring with an inclination as high as 66$\degree$ would
be inconsistent with the narrow CO line and compact size from
interferometry. To reconcile this difference, the system could contain a
compact gas disc or ring with low inclination, but with
a highly flared dust disc; alternatively the gas lies in a more
extended ring, but the line brightness is low because of CO
depletion and excitation conditions in the disc atmosphere, which have
not been included in the present model.

\subsubsection{HD135344}

This is the one of the closest gas-rich discs to the Sun (CWD98), and
also the oldest known star with such a disc:
Thi et al. (2001) derive an age of 16.7 Myr.
No scattered light has been detected using coronograph observations
down to radii of $\sim 1$ arcsec although a binary pair was found at
separation $\sim 5.8$ arcsec (490 au) from the primary (Augereau et al., 2001).
The CO line is double-peaked (Fig.~2b) and can be well fitted by a
compact disc of outer radius 75 au viewed nearly face-on.  A more extended
edge-on ring is excluded by the presence of wings in the line profile.
If we assume the
nearby binary pair are associated with HD135344, this would imply an upper
limit of $\sim$200~au to the disc radius, as material is unlikely
to exist in a stable orbit at radii within a factor of $\sim$2 of the 
binary separation.

\subsubsection{HD141569}

An extended region of scattered light from a circumstellar
disc has been imaged by several authors using coronography
(e.g. Mouillet et al., 2001, and refs. therein), and both
this and mid-infrared images indicate an inclination of 52$\degree$.
In addition,
CO emission was detected from this object by Zuckerman et al. (1995),
and recently imaged using interferometry (Augereau et al., 2004).
We obtained a deep CO spectrum (Fig. 2b), which
shows a double-peaked line profile, with a distinctive ``shoulder''
at higher relative velocities. A single ring or disc structure
cannot explain this lineshape, giving for example, the model
shown by the dashed line in Fig.~2b. Instead we adopted a double ring
structure. The inclination of the dust disc is well
constrained by the images so we adopt the same value for the gas.
The relatively narrow overall line width indicates ring radii of 95
and $\sim$250~au (shown by the solid line on Fig.2b).
The latter is similar to the radius found in the recent
interferometry results.
The low disc mass and high UV luminosity of the star
requires the rings to be physically thin in order to maintain high
extinction and avoid photodissociating the CO. The
best fit to the data has an opening angle $\theta = 1.0\degree$,
consistent with the low fractional infrared excess of the star
(see Table~1). The low
observed CO brightness temperature implies the surface area, and therefore the
radial extent of the rings, must also be small. In order to fit the
intensities, radial widths of $\sim$20~au and 8~au are required for the
inner and outer rings.

The dust rings seen in scattered light reach
peak densities at radii $\sim$200 and $\sim$350 au (Mouillet et al.). The
gaps detected in dust extends from
$\leq$125 - 175 au (measurement of the inner radius was limited
by the coronograph occulting bar) and 215 - 300 au. Our CO model
suggests emission also arises
from a ring inside this, at radius 90~au, similar
to the size of the mid-infrared source (Fisher et al., 2000).
Brittain et al. (2003) also detected near-infrared ro-vibrational CO
lines from the disc; the linewidths and temperatures indicate that this
arises at radii 17 - 50~au, which may correspond
with the innermost edge of our inner CO ring.

A further feature of the spectrum in Fig.~2b is the assymmetry of
the narrow component; at positive velocities the line is $\sim50$ per cent
brighter than at negative velocities. As noted for MWC480 above, if the CO
emission is optically thick, this would require a large variation in
density within the ring (or a smaller variation in temperature). In
the case of HD141569, a large variation in CO density could be explained
by a difference in the extinction to the star (and hence differences
in photodissocation rate). A recent K-band image
(Boccaletti et al., 2003) also shows an assymmetry in scattered light,
whereby the north-eastern sector of the 200~au ring is significantly brighter.
If this is related to the CO assymmetry, it would indicate this is the
side of the ring approaching us.

Overall the CO model of HD141569 suggests that molecular gas and dust
is well intermixed in a physically thin pair of rings.
A more detailed comparison between the scattering dust and molecular
gas would require a high resolution CO map of this complex system.

\subsubsection{HD163296}

The disc around this star has been resolved both in scattered light and
mm-wave interferometry (Mannings \& Sargent, 1997; Grady et al., 2000).
Our high s:n spectrum shows reasonable agreement with a model disc inclined at
$30\degree$ (Fig.~2c), although the deep dip near the stellar systemic velocity
cannot be reproduced in a simple model. One possible explanation for
this might be self-absoption from cool foreground gas, although
the low optical extinction to the star suggests little material exists along
the line of sight (van den Ancker et al., 1998).
The CO spectra of Thi et al. (2004) shows line-of-sight gas
at velocities considerably different from that of the star (and outside the
range shown in Fig.~2c); however, there is no evidence of extended
emission at the stellar velocity itself.
The derived disc inclination is inconsistent with that
found from the morphology
of the interferometric map - we cannot reproduce the line profile with
inclinations as high as 60$\degree$.
However, the derived outer radius is in close agreement with the
interferometric result, and is similar to the inner edge
of the dark lane seen in scattered light, indicating that
the gas and dust co-exist within the disc.
Modelling the CO emission profile out to relative velocities
of $\pm 4$km s$^{-1}$ suggests that gas exists in the disc down
to radii less than 20~au. No additional high-velocity molecular outflow
component is needed to explain the line shape.

\subsubsection{UX Ori}

Extended CO emission is seen in this region at v$\sim27$km s$^{-1}$
in both the signal and reference positions,
however, a narrow emission line was also seen at the stellar velocity
itself (Fig.~2d).
All model fits which assume the emission is from a disc require an orientation 
close to face-on ($i \leq 20\degree$) to reproduce the line width.
This is inconsistent with the interpretation of the variability as
obscuration in a near edge-on disc (Natta et al., 1999).
The line may instead be from ambient gas more distant from the star,
possibly associated with the extended far-infrared emission seen by Natta
et al.; mapping of this emission would help identify its origin.

\subsubsection{TW Hya}

We obtained a deep CO spectrum of disc around
the 8~Myr old star TW Hya, known to
have numerous emission lines at sub-mm wavelengths (e.g. van Zadelhoff et al.,
2001). Although TW Hya is a K7 rather than an AeBe star, the disc is well
known to be close to face-on; we have included it in this study
partly to compare the results from our simple
LTE model with the more detailed radiative transfer and chemistry
modelling carried out by
van Zadelhoff et al., and partly to search for weak higher-velocity gas
in the deep spectrum.
Assuming standard $^{12}$CO abundance, we derive a disc outer radius of 160 au,
very similar to the scattered light images, as well as the size derived
from the SED (Calvet et al., 2002) and that adopted by
van Zadelhoff et al. Furthermore the CO
abundance depletions measured by van Zadelhoff et al.
(up to 150) do not significantly affect the derived
result; for such a depletion, the disc size required to fit the present data
increases only slightly (190 au), confirming that
the emission is indeed extremely optically thick.

The narrow line width seen
in Fig.~2d indicates an inclination of only $i\sim5\degree$ and a
physically thin disc ($\theta \sim 5\degree$). However, weak
wings reaching to $\pm 1.3$km s$^{-1}$ can be seen in the deep spectrum.
These require a model where the gas disc extends in
to a radius $\leq$30 au (for comparison, a model with a central hole of
radius 50 au gives the spectrum shown by the dashed line).
This compares with the innermost edge of the dust disc and the
developing gap (4 au) derived by Calvet et al. from the continuum SED.

\section{Discussion}

\subsection{Disc outer radii}

The outer disc radii derived in our sample of stars
mostly lie between 30 and 400 au, with a mean of
$\sim$150 au. Most are larger than the solar system ($\sim$50 au to the edge
of the Kuiper Belt), and are closer to the size of
the debris discs seen around older stars (typically 50 - 150 au,
e.g. Holland et al., 1998). However, as many as 5 of the gas discs
are compact (R$_{out} \leq$50 au), and at least another 7 have evidence 
of gas at inner radii smaller than 50 au.
So what determines the radius, and how is the gas removed?

Figure~3 shows that the disc size is not correlated with the
stellar spectral type, with a Pearson correlation coefficient of -0.04.
This applies over 2.5 orders of magnitude in
luminosity. We also include published T Tauri disc radii from Simon
et al. (2000) as well as model fits to the CO upper limits for stars
in Table~1 with $f \geq 0.1$, ie which are predicted to have measurable
CO emission (see discussion above).

Figure~3 also shows the radius at which a black body in thermal
equilibrium with the central star reaches the
CO freeze-out temperature (assumed to be 20K). This
assumption of black body temperature would require large grain
sizes ($> 1mm$), as well as low extinction to the star.
It is likely that the disc mid-plane temperature would
be lower than this in some objects due to the high extinction.
Conversely, if the grain sizes were closer to those in the
interstellar medium (ie
little grain growth had occured) the temperature would be higher.
If freeze-out were important, we would not expect to see disc radii
above this line. Fig.~3 shows that this
is indeed the case for all stars of type F or earlier. However, some
T Tauri stars lie above this line, suggesting that another source of
energy such as accretion luminosity may be heating these discs.

\epsfverbosetrue
\epsfxsize=8.0 cm
\begin{figure}
\center{
\leavevmode
\epsfbox{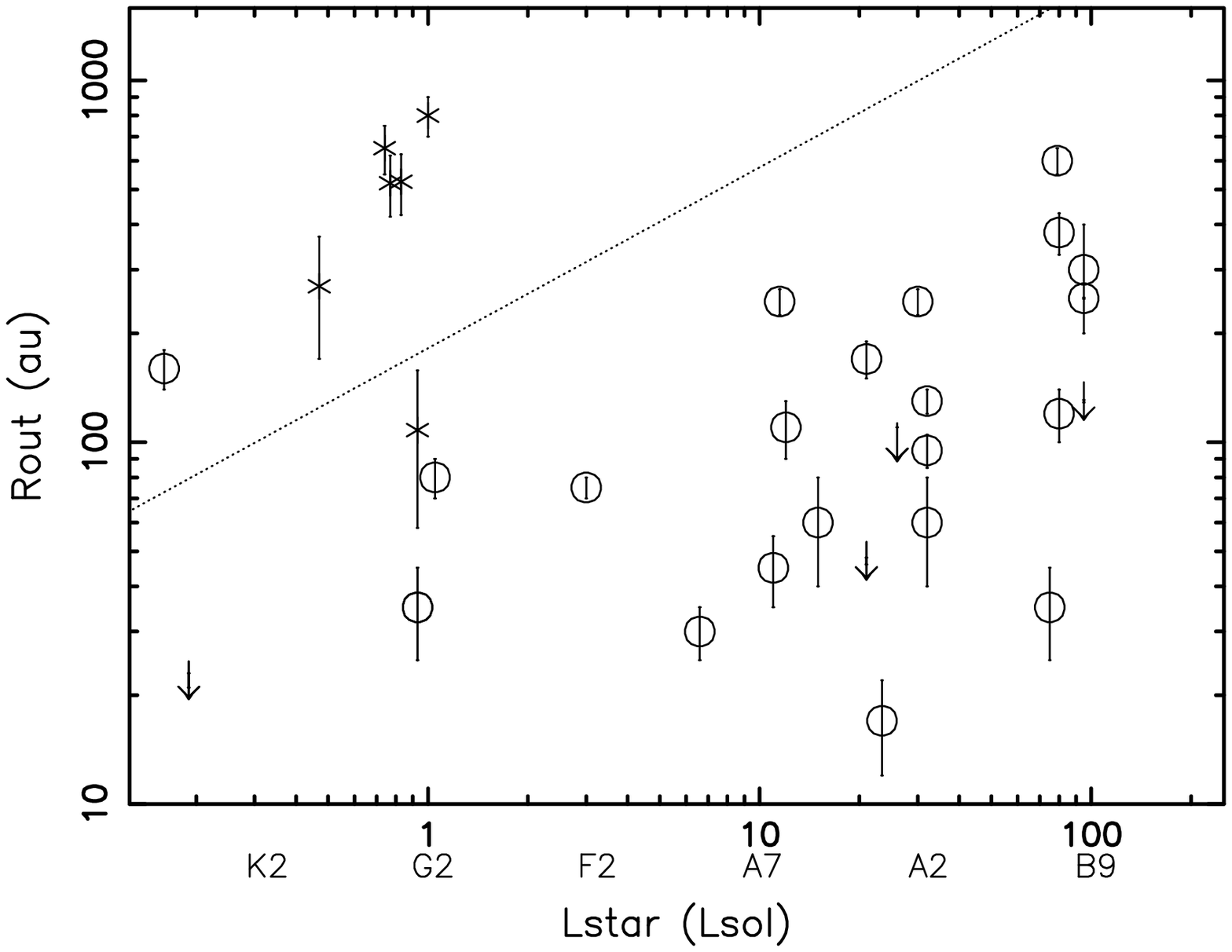}
\caption {
Derived outer radii of gas disc plotted as a function of stellar luminosity
and Zero Age Main Sequence spectral type.
The straight line shows the temperature at which CO freezes out onto grains
(see text). Also shown
are derived upper limits (2-$\sigma$) to the disc size for stars 
with no line detection, but which have $f \geq 0.1$.
The star symbols represent the interferometric observations
of discs around T Tauri stars from Simon et al. (2000).
}
}
\end{figure}

\subsection{Comparison with dust discs}

The continuum SEDs from many of the objects in Table~2 have been
used to derive disc sizes, based on standard dust models (e.g. Dominik et
al., 2003). In all cases except HD135344, the SED-derived radii are
smaller than those derived
from CO; we attribute this to the greater sensitivity of the J=3-2
transition of CO to a given mass of material. However, it is possible
that large grains could accrete rapidly through a disc, leaving
a gas disc larger than the dust disc (Takeuchi \& Lin, 2002).

We can also compare the total disc mass (derived from the continuum
SED, scaled assuming a constant gas:dust ratio of 100) with the gas
disc outer radius (Fig.~4). There is some evidence of a
correlation between the disc size
and mass, with a Pearson correlation coefficient of 0.62. The slope
of the fit is 2.8, close to a value of 3 which would apply if the mean density
in the discs was constant. Further data over
a wider range of masses would be needed to confirm this result.

\epsfverbosetrue
\epsfxsize=8.0 cm
\begin{figure}
\center{
\leavevmode
\epsfbox{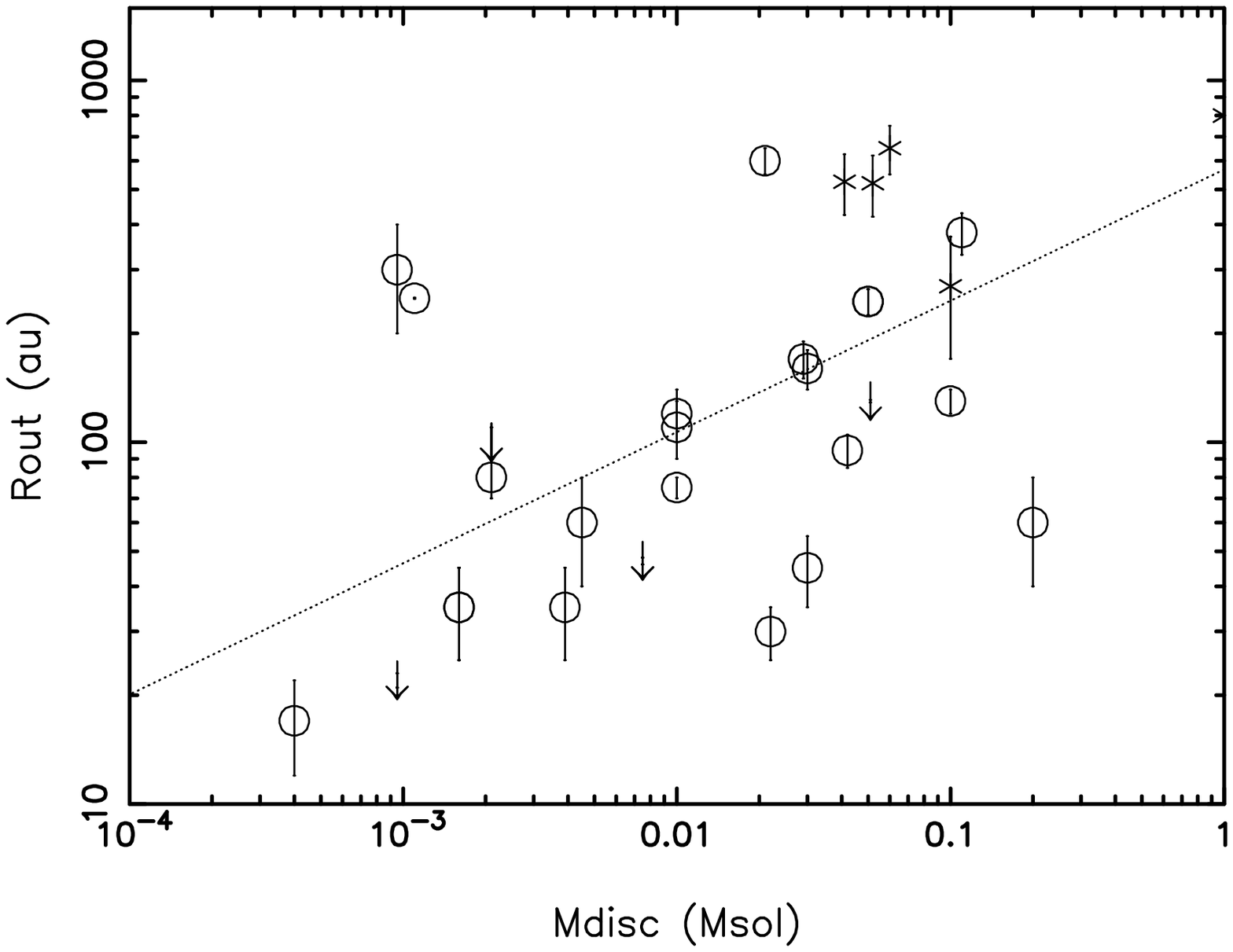}
\caption {
Derived outer radii of gas discs plotted as a function of disc mass
from dust continuum fluxes.
Disc mass includes the gas, assuming a constant gas:dust ratio of 100.
The upper limits to the sizes represent stars in the present survey with
$f \geq 0.1$.
The star symbols represent the interferometric observations of discs around
isolated T Tauri stars from Simon et al. (2000). The line is a fit to the
data, with a slope of 2.8.
}
}
\end{figure}

Scattered light from dust around four discs in Table~2 has been
detected using coronography (AB Aur, HD141569, HD163296 and TW Hya). 
These are all predicted from the CO modelling to have angular sizes
$> 3$ arcsec. Their optical sizes
and distributions are similar to those derived from CO, implying
that the gas and small scattering dust grains
are closely intermixed. The predicted angular size of MWC480
also suggests that scattered light may be detectable, however,
coronograph imaging has not shown
extended emission (Augureau et al., 2001).
One possible explanation is that this disc is not flared, or
is shadowed by a thick inner disc.

\subsection{Gas disc lifetimes}

Many authors have used disc masses derived from mm-wavelength
continuum or line fluxes to estimate the disc lifetime
(e.g. Beckwith et al., 1990; Thi et al., 2001), and the 
results suggest that dust discs survive for typically $\sim 10^7$ yr.
However, mass estimates are uncertain because the detectable dust
is part of a broad grain size distribution and so represents only a small
fraction of the total solid mass. The gas:dust ratio is
also uncertain (e.g. Thi et al., 2001), and is likely to evolve in these discs.
Rather surprisingly, the
results show no evidence of evolution of disc {\em mass}
between ages of 10$^5$ and 10$^7$ yr, around either
T Tauri (Beckwith et al., 1990) or HAeBe stars (Natta et al., 2000).

Although we cannot derive gas masses from $^{12}$CO (see above), we can use
the results to see whether there is an evolution in {\em size} of the gas
disc; this is illustrated in Figure~5. Typical uncertainties in the age
derivations are 1-3 Myr, and dominate the plot. However, splitting the
sample in two based on their age, the discs around older objects
(7-20 Myr) have a mean radius of 75~au, significantly
smaller than the younger discs (3-7 Myr), where the mean radius is
210~au. There are no large, old gas discs (which would appear in the top right
of the figure). This effect is
strengthened further if the published results from
relatively young (1-5 Myr) T Tauri discs are included. Although
these have significantly lower luminosity, the lack of
dependence of size with spectral type (see Fig.~3) indicates that 
disc age is more important than stellar luminosity.
Discs around even younger stars (Class I or II YSOs, typically of
age 0.1-1 Myr) are difficult
to separate from surrounding ambient clouds, both
observationally and in the definition of where the disc ends and
``ambient'' gas begins. Fuente et al. (2002) found that the very
large-scale ($\geq 10^4$au) ambient
gas around HAeBe stars disappears by $\sim 1$Myr.
However interferometric observations show discs with radii of a
few $10^2$ to $\sim 10^3$au around some Class I stars,
thought to be $\leq 1$Myr in age (e.g. Mundy et al., 2000). 
Overall Fig.~5 shows evidence for evolution of the gas disc size, such
that R$_{out}$ decreases by a factor of $\sim$3 over the 3-20 Myr period.
An alternative view is that the lifetime of the outer $\sim$200au
gas disc
region is $\sim 7$Myr, and the inner $\sim$75au region is $\sim$20Myr.
However, this result should be treated with caution, as the age
determinations are somewhat uncertain.

\epsfverbosetrue
\epsfxsize=8.0 cm
\begin{figure}
\center{
\leavevmode
\epsfbox{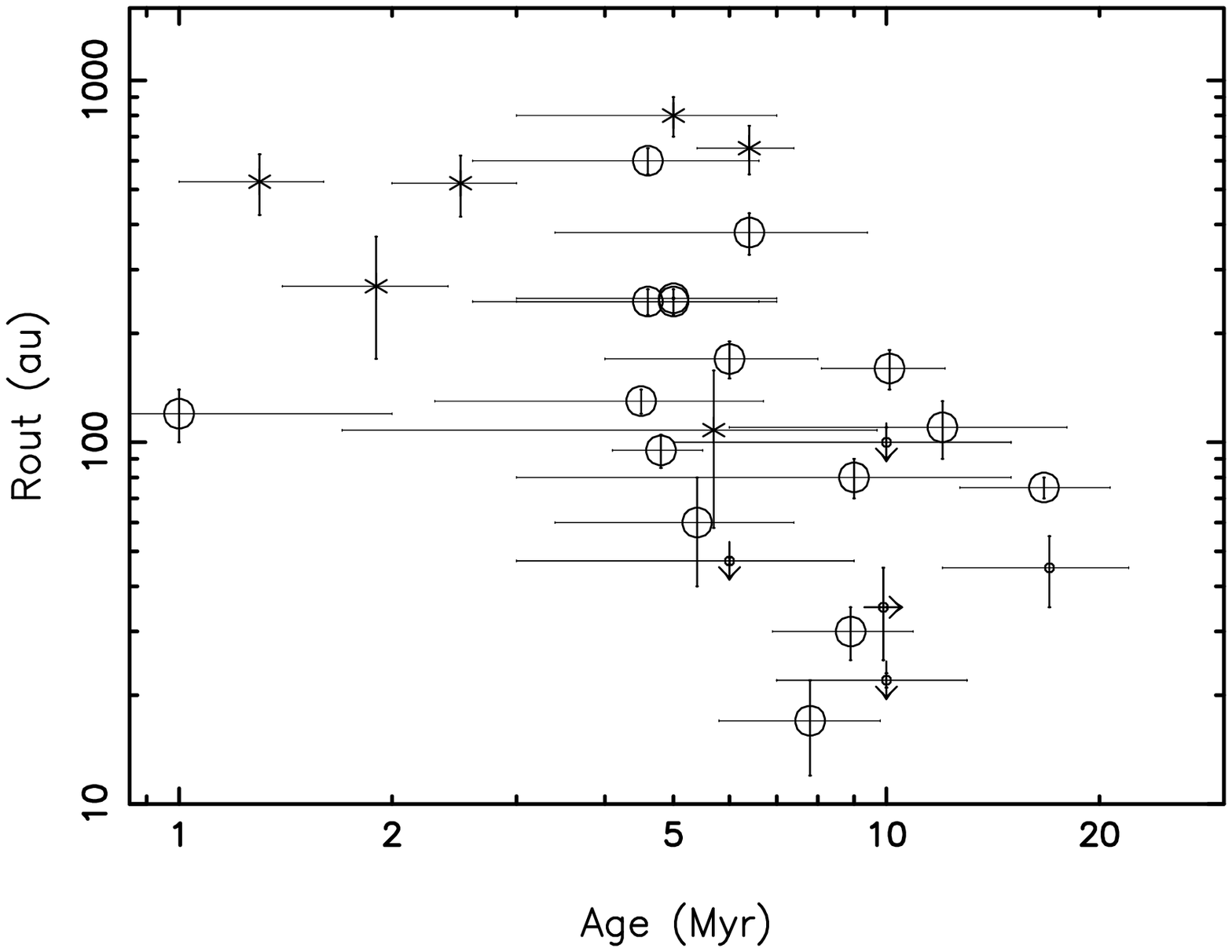}
\caption {
Derived outer radii of gas discs plotted as a function of stellar age.
Large circles represent stars whose ages are well-determined using stellar
evolutionary tracks; smaller circles represent stars where the age is
more uncertain. Ages are taken from Zuckerman et al (1995),
Natta et al. (1997), Thi et al. (2001), van den Ancker et al. (1998),
Dominik et al (2003) and Natta et al. (2004). The upper limits to the sizes
represent stars in the present survey with
$f \geq 0.1$ which have known ages.
The star symbols represent the interferometric observations of discs around
isolated T Tauri stars from Simon et al. (2000).
}
}
\end{figure}

Models of disc removal suggest that timescales should {\em increase} with
radius (e.g. Hollenbach et al., 2000). So we might expect
relatively little evolution of the outer radius except
at the end of the disc lifetime, when the gas rapidly disappears
(Clarke et al., 2002) or the obscuring dust
coagulates, allowing photodissociation of the outer regions. However, the
dominant removal
mechanism in the outer (100-1000 au) region of the disc may be external
photoevaporation (Matsuyama et al., 2003) or interaction with other
stars in a young cluster (Mundy et al., 2000).
The decrease in mean outer radius in Fig.~5 would therefore
be determined by interaction with the local environment, rather than
evolution of the star itself.

Finally the results of the modelling indicate that at
least 10$^{21}$kg of gaseous CO exists for $\sim$20 Myr at
radii $\leq 100$au around these stars. This could be available to
form $\sim 10^9$ 1km-sized comets, assuming a cometary CO
mass fraction of 10 per cent.

\section{Conclusions}

We have surveyed a sample of 59 isolated HAeBe stars and Vega-excess
stars for sub-mm $^{12}$CO emission. A clear correlation with fractional
excess is seen, with a high CO detection rate around stars with $f \geq 0.1$
(18/27) and a low detection rate for those with $f < 0.01$ (2/27).
This is explainable by CO dissociation by stellar UV photons; the stars
showing high $f$ have an optically thick disc shielding the gas.
Most discs with detected CO have derived
opening angles $\geq 12\degree$, although
in the low-$f$ objects, the shielding disc must be geometrically thin
($\theta \leq 1\degree$). The data are consistent with {\em all}
of the $f > 0.1$ discs
having significant gas (with a minimum mass of $10^{-3}$M$_J$),
suggesting that CO will exist as long as it is sufficiently
shielded from stellar UV.

Approximately 50 per cent of the CO lines are double-peaked, which we
interpret as arising in a Keplerian disc. Comparing a basic model with the CO
profiles, we derive outer radii
of 20-300 au for the majority of the discs. In most cases these radii are
considerably greater than those derived
from continuum SED modelling. These gas-rich isolated
discs are found around stars of
ages from 3-17 Myr, and spectral types K through to B9. There is no dependence
of size with the spectral type. But there is evidence of a correlation
between radius and disc mass, with $M_d \propto R_{out}^{2.8}$.
Also the outer radius is dependent on disc age: splitting the stars into
two age groups, we find stars $<$ 7 Myr in age have discs 3 times larger
than those of age 7-20~Myr. The results indicate that gas-rich discs are
available to form planets and cometary material for
$\sim 10^7$yr at radii $\sim$200~au, and may last a factor of
two longer in the inner $\sim$75~au region.

In the four largest CO discs, published scattered light images show
comparable dust and gas outer radii, suggesting that the scattering dust and
molecular gas are co-located within these discs.
In only one case (HD141569) is there evidence of a central gas-free region;
many of the others show evidence that gas is found down to
radii smaller than 30au.

\section*{Acknowledgments}

The James Clerk Maxwell Telescope is operated by the Joint Astronomy
Centre on behalf of the United Kingdom Particle Physics and Astronomy
Research Council, the Netherlands Organisation for Scientific Research,
and the National Research Council of Canada. We wish to thank the various
observers, particularly the local staff in Hawaii, for staying at the
telescope during mediocre weather, allowing the backup programs to
be carried out. Some of the data was extracted from the JCMT
archive at the Canadian Astronomy
Data Center, which is operated by the Dominion Astrophyical Observatory for
the National Research Council of Canada's Herzberg Institute of Astrophysics.

\section{References}

\parindent 0mm

Aikawa, Y., Herbst, E., 2001, \aa, 371, 1107

Augereau, J. C., Lagrange, A. M., Mouillet, D., M\'enard, F.,
2001, \aa, 365, 78

Augereau, J.C., Dutrey, A., Lagrange, A. M., Forveille, T., 2004,
\aa (submitted)

Aumann, H.H., Beichman, C. A., Gillett, F. C., de Jong, T., Houck, J. R.,
Low, F. J., Neugebauer, G., Walker, R. G., Wesselius, P. R., 1984, \apj,
278, L24

Backman, D.E., Paresce, F., 1993, in Levy, E.H., \& Lunine. J.I., eds,
Protostars and Planets III, Univ. of Arizona, Tucson, p. 1253

Beckwith, S.V.W., Sargent, A.I., Chini, R.S., Guesten, R., 1990,
\aj, 99, 924

Beckwith, S.V.W., Sargent, A.I., 1993, \apj, 402, 280

Beskrovnaya, N.G., Pogodin, M.A., Miroshnichenko, A.S., Th\'e, P.S.,
Savanov, I.S., Shakhovskoy, N.M., Rostopchina, A.N., Kozlova, O.V.,
Kuratov, K.S., 1999, 343, 163

Boccaletti, A., Augereau, J.-C., Marchis, F., Hahn, J., 2003, \apj, 585, 494

Brittain, S.D., Rettig, T.W., Simon, T., Kulesa, C., DiSanti, M.A.,
Dello Russo, N., 2003, \apj, 588, 535

Calvet, N., D'Alessio, P., Hartman, L., Wilner D., Wlash, A., Sitko, M., 2002,
\apj, 568, 1008

Coulson, I.M., Walther, D.M., Dent, W.R.F., 1998, \mnras, 296, 934

Dent, W.R.F., Greaves, J.S., Mannings, V., Coulson, I.M., Walther, D. M.,
1995, \mnras, 277, L25

Dent, W.R.F., Walker, H.J., Holland, W.S., Greaves, J.S., 2000, \mnras,
314, 702

Dominik, C., Dullemond, C.P., Waters, L.B.F.M., Walch, S., 2003,
\aa, 398, 607

Dunkin, S.K, Barlow, M.J., Ryan, S.G., 1997, \mnras, 290, 165

Dunkin, S.K., Crawford, I.A., 1998, \mnras, 298, 275

Fekel, F.C., Webb, R.A., White, R.J., Zuckerman, B.,
1996, \apj, 462, L95

Fisher, R.S., Telesco, C.M., Pina, R.K., Knacke, R.F., Wyatt, M.C.,
2000, \apj, 532, L141

Fuente, A., Martin-Pintado, J., Bachiller, R., Neri, R., Palla, F., 1998,
\aa, 334, 253

Fuente, A., Martin-Pintado, J., Bachiller, R., Rodriguez-Franco, A.,
Palla, F., 2002, \aa, 387, 977

Grady, C.A., Woodgate, B., Bruhweiler, F.C., Boggess, A., Plait, P.,
Lindler, D.J., Clampin, M., Kalas, P., 1999, \apj, 523, L151

Grady, C. A., Devine, D., Woodgate, B., Kimble, R., Bruhweiler, F. C.,
Boggess, A., Linsky, J. L., Plait, P., Clampin, M., Kalas, P.,
2000, \apj, 544, 895

Greaves, J.S., Mannings, V., Holland, W.S., 2000, Icarus, 143, 155

Heap, S.R., Lindler, D.J., Lanz, T.M., Cornett, R.H., Hubeny, I.,
Maran, S. P., Woodgate, B., 2000, \apj, 539, 435

Herbig, G.H., 1960, \apjs, 4, 337

Holland, W. S., Greaves, J. S., Zuckerman, B., Webb, R. A., McCarthy, C.,
Coulson, I. M., Walther, D. M., Dent, W. R. F., Gear, W. K., Robson, I., 1998,
Nature, 392, 788

Hollenbach, D.J., Yorke, H.W., Johnstone, D., 2000,
in Mannings, V., Boss, A., Russell. S.S., eds, Protostars and Planets IV,
Univ. of Arizona, Tucson, p. 401

Hollenbach, D.J., Takahashi, T., Tielens, A.G.G.M., 1991,
\apj, 377, 192

Jayawardhana, R., Fisher, R.S., Telesco, C.M., Pina, R.K.,
Barrado y Navascu\'es, D. Hartmann, L.W., Fazio, G.G., 2001, \aj
122, 2047

Kalas, P., Graham, J.R., Beckwith, S.V.W., Jewitt, D.C., Lloyd, J.,
2002, \apj, 567, 999

Kamp, I., van Zadelhoff, G.-J., van Dishoeck, E.F., Stark, R., 2003,
\aa, 397, 1127

Kastner, J.H., Zuckerman, B., Weintraub, D.A., Forveille, T., 1997, Science,
277, 67

Lisse, C., Schultz, A., Fernandez, Y., Peschke, S., Ressler, M.,
Gorjian, V., Djorgovski, S.G., Christian, D.J., Golisch, B., Kaminski, C.,
2002, \apj, 570, 779

Li, A., Lunine, J.I., 2003, \apj, 594, L987

Liseau, R., 1999, \aa, 348, 133

Malfait, K., Bogaert, E., Waelkens, C., 1998, \aa, 331, 211

Mannings, V., Barlow, M.J., 1998, \apj, 497, 330

Mannings, V., Sargent, A.I., 1997, \apj, 490, 792

Mannings, V., Koerner, D. W., Sargent, A. I., 1997, Nature, 388, 555

Mannings, V., Sargent, A.I., 2000, \apj, 529, 391

Matsuyama, I., Johnstone, D., Hartman, L., 2003, \apj, 582, 893

Meeus, G., Waters, L.B.F.M., Bouwman, J., van den Ancker, M.E.,
Waelkens, C., Malfait, K., 2001, \aa, 365, 476

Mouillet, D., Lagrange, A.M., Augereau, J.C., M\'enard, F., 2001, \aa, 
372, L61

Mundy, L.G., Looney, L.W., Welch, W.J., 2000, in Mannings, V., Boss, A.,
Russell, S.S., Protostars and Planets IV, Univ. of Arizona, Tucson, p. 355

Natta, A., Testi, L., Neri, R., Shepherd, D.S., Wilner, D.J., 2004,
\aa, 416, 179

Natta, A., Whitney, B. A., 2000, \aa, 364, 633

Natta, A., Grinin, V. P., Mannings, V., Ungerechts, H., 1997, \apj, 491, 885

Natta, A., Prusti, T., Neri, R., Thi, W. F., Grinin, V. P., Mannings, V.,
1999, \aa, 350, 541

Natta, A., Grinin, V.P., Mannings, V., 2000, in Mannings, V., Boss, A.,
Russell, S.S., eds, Protostars and Planets IV, Univ. of Arizona, Tucson, p. 559

Natta, A., Prusti, T., Neri, R., Wooden, D., Grinin, V.P., Mannings, V.,
2001, \aa, 371, 186

Omodaka, T., Kitamura, Y., Kawazoe, E., 1992, \apj, 396, L87

Pi\'etu, V., Dutrey, A., Kahane, C., 2003, \aa, 398, 565

Simon, M., Dutrey, A., Guilloteau, S., 2001, \apj, 545, 1034

Sylvester, R.J., Skinner, C.J., Barlow, M.J., Mannings, V., 1996, \mnras, 279,
915

Sylvester, R.J., Dunkin, S.K., Barlow, M.J., 2001, \mnras, 327, 133

Sylvester, R.J., Mannings, V., 2000, \mnras, 313, 73

Takeuchi, T., Lin, D.N.C., 2001, \apj, 581, 1344

Testi, L., Natta, A., Shepherd, D.S., Wilner, D.J., 2003, \aa, 403, 323

Th\'e, P.S., de Winter, D., Perez, M.R., 1994, \aas, 104, 315

Thi, W.F., van Dishoeck, E.F., Blake, G.A., van Zadelhoff, G.J.,
Horn, J., Becklin, E.E., Mannings, V., Sargent, A.I., van den Ancker,
M. E., Natta, A., Kessler, J., 2001, \apj, 561, 1074

Thi, W.F., van Zadelhoff, G.-J., van Dishoeck, E.F., 2004, \aa, 425, 955

Torres, G., 2004, \aj, 127, 1187

van den Ancker, M.E., de Winter, D., Tjin A Djie, H.R.E., 1998, \aa, 330, 145

van Zadelhoff, G.-J., van Dishoeck, E.F., Thi, W.-F., Blake, G.A.,
2001, \aa, 377, 566

van Zadelhoff, G.-J., Aikawa, Y., Hogerheijde, M. R., van Dishoeck, E. F.,
2003, \aa, 397, 789

Yamashita, T., Handa, T., Omodaka, T., Kitamura, Y., Kawazoe, E.,
Hayashi, S.S., Kaifu, N., 1993, \apj, 402, L65

Zuckerman, B., Forveille, T., Kastner, J.H., 1995, Nature, 373, 494

\end{document}